\documentclass[%
 aip,
 amsmath,amssymb,
 reprint,%
]{revtex4-1}
\usepackage{color,hyperref,amsmath}
\usepackage{graphicx}% Include figure files
\usepackage{dcolumn}% Align table columns on decimal point
\usepackage{bm}% bold math
\usepackage[utf8]{inputenc}
\usepackage[T1]{fontenc}
\usepackage{mathptmx}
\usepackage{etoolbox}

\makeatletter
\def\@email#1#2{%
 \endgroup
 \patchcmd{\titleblock@produce}
  {\frontmatter@RRAPformat}
  {\frontmatter@RRAPformat{\produce@RRAP{*#1\href{mailto:#2}{#2}}}\frontmatter@RRAPformat}
  {}{}
}% 
\makeatother 
\begin{document}
\title{Exploration of field-like torque and field-angle tunability in coupled spin-torque nano oscillators for synchronization}
\author{R.~Arun$^{1}$}
\author{R.~Gopal$^{2}$}
\author{V.~K.~Chandrasekar$^{2}$}
\author{M.~Lakshmanan$^1$}
 
\affiliation
{
$^{1}$Department of Nonlinear Dynamics, School of Physics, Bharathidasan University, Tiruchirapalli-620024, India\\
$^{2}$Department of Physics, Centre for Nonlinear Science \& Engineering, School of Electrical \& Electronics Engineering, SASTRA Deemed University, Thanjavur- 613 401, India. \\
}
\date{\today}

\date{\today}
\begin{abstract}
We investigate the influence of field-like torque and the direction of the external magnetic field on a one-dimensional array of serially connected spin-torque nano oscillators, having free layers with perpendicular anisotropy, to achieve complete synchronization between them by analyzing the associated Landau-Lifshitz-Gilbert-Slonczewski equation. The obtained results for synchronization are discussed for the cases of 2, 10 and 100 oscillators separately. The roles of the field-like torque and the direction of the external field on the synchronization of the STNOs are explored through the Kuramoto order parameter. While the field-like torque alone is sufficient to bring out global synchronization in the system made up of a small number of STNOs, the direction of the external field is also needed to be slightly tuned to synchronize the one-dimensional array of a large number of STNOs. The formation of complete synchronization through the construction of clusters within the system is identified for the 100 oscillators. The large amplitude synchronized oscillations are obtained for small to large numbers of oscillators. Moreover, the tunability in frequency for a wide range of currents is shown for the synchronized oscillations up to 100 spin-torque oscillators. In addition to achieving synchronization, the field-like torque increases the frequency of the synchronized oscillations. The transverse Lyapunov exponents are deduced to confirm the stable synchronization in coupled STNOs due to the field-like torque and to validate the results obtained in the numerical simulations. The output power of the array is estimated to be enhanced substantially due to complete synchronization by the combined effect of field-like torque and tunability of the field angle.
\end{abstract}

%\pacs{ 05.45.-a, 05.45.Xt, 89.75.-k}

\maketitle
\textcolor{blue}{\bf \it This article is in honor of Prof. J$\ddot{u}$rgen Kurths on his seventieth birthday, wishing him many more active years of academic life.}

\begin{quotation}
	Since the last decade, the study of spin-torque nano oscillators (STNOs) has attracted considerable interest. They have been considered not only as magnetic models to study nonlinear dynamics at the nanoscale level but also as promising candidates as nanoscale microwave generators, which show many promising applications. Although the STNO can present such an excellent frequency tunability in the GHz range and is fully compatible with a high level of integration, two significant constraints are faced by an STNO in achieving the required high output power and low linewidth. One of the solutions to gain the needed output power is by synchronizing an array of STNOs through stable synchronized oscillations. Due to their highly nonlinear behaviour and wideband frequency tunability, STNOs can be synchronized by taking advantage of different types of coupling. There has been a few attempts in the literature to achieve synchronization of STNOs in large arrays through appropriate interactions and couplings. In this article, we propose to investigate an one-dimensional array of STNOs connected in series, which couple themselves through self-emitted currents under field-like torque for their synchronization. The significance role of field-angle (angle between external magnetic field and out-of-plane axis) is carefully explored. Our main objective here is a demonstration of approaching a global synchronization in an extensive array of STNOs through different collective dynamical modes. Our findings may have potential application in STNO-based information processing and digital computation applications.
\end{quotation}

\section{Introduction}

The study of synchronization phenomenon in STNOs has been the subject of active research in recent years due to its potential applications for the generation of microwave signals in the nanoscale regime ~\cite{man,gro,uraz,kend,tur}. Some significant efforts have been made to study magnetization dynamics and synchronization of coupled STNOs driven by spin-polarized current~\cite{slon:96}, external microwave current or through a microwave magnetic field~\cite{adler}, spin waves~\cite{kend}, magnetic fields~\cite{subash1,subash2,gopal} and electrical couplings~\cite{rippard:05}. The process of synchronization of STNOs is more desirable for enhancing the efficiency, the quality factor and the oscillation frequency of the practical STNO devices such as the high-density microwave signal processors and chip-to-chip communication systems~\cite{rippard:05,geor,zhou,nakada,zeng:12}. Further, it has attracted much attention from the viewpoints of fundamental physics and practical applications such as brain-inspired computing and microwave-assisted magnetic reading~\cite{chio:14,loca:14,gro1,kudo}.

Studies of $N$ electrically coupled STNOs show that each STNO leads to act as feedback between the STNOs, causing them to synchronize, and collectively the power of the  microwave  output of the array increases by $N^{2}$ times~\cite{tur1}. A small region of synchronization of serially connected STNOs is identified in the parameter space~\cite{per}, and the inadequacy of the synchronization region is due to the coexistence of multiple stable attractors which leads to the synchronization regime being susceptible to initial conditions~\cite{li}. Turtle \emph{et} al showed that the basin of attraction for getting the synchronized oscillations enhances by changing the angle of the applied magnetization field~\cite{tur}. Further, the existence and stability of the synchronized state and the conditions to synchronize the individual precessions have also been studied in an array of $N$ serially connected identical STNOs coupled through current, which have been clearly demonstrated in Ref~\cite{tur}. Recently, mutual synchronization between two parallelly connected STNOs, coupled by current, has also been identified~\cite{tomo:18}.  

However, the significant issues in the system of coupled STNOs include the formation of multistability and decrement in frequency while increasing the current, especially at larger currents~\cite{tur}. The existence of multistability prevents the system from exhibiting stable synchronized oscillations for all initial conditions. Removing this multistability and making the system exhibit stable synchronized oscillations are challenging tasks. Also, a decrease in the frequency of synchronized oscillations while increasing the current limits the enhancement of frequency beyond particular value.  

\begin{figure*}[htp]
	\centering\includegraphics[angle=0,width=0.7\linewidth]{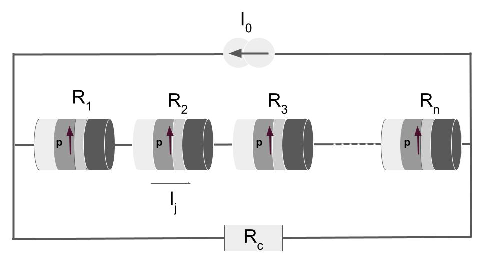}
	\caption{1-D array of $N$ STNOs electrically coupled in series with all-to-all coupling. $R_j$ and $I_j$ are the resistance and current passing through the $j^{th}$ STNO. $R_C$ is the load resistance, $I_0$ is the input current and ${\bf p}$ is the unit magnetization vector of the pinned layers.}
	\label{fig1}
\end{figure*}

These two important issues are addressed in this paper by investigating the global synchronization in serially coupled STNOs having perpendicular magnetic anisotropy for their free layers.  Moreover, the global synchronization for the STNOs with perpendicular anisotropy is yet to be studied. A recent work by the present authors confirms that the multistability can be removed by incorporating the field-like torque to a system of coupled STNOs \cite{arun_ieee}. In addition to this, the field-like torque is capable of stabilizing self oscillations in STNOs \cite{tani_apl} and spin-Hall oscillators \cite{arun_jpcm}.  In this connection we are motivated to investigate the role of field-like torque on synchronizing serially connected and electrically coupled STNOs having free layers with perpendicular magnetic anisotropy and in-plane magnetized pinned layers.  The main reason for choosing perpendicular magnetic anisotropy is that it is robust against thermal fluctuations and provides high storage density \cite{tudu,sbiaa}.  Also, this type of STNO with perpendicularly magnetized free layer and in-plane magnetized pinned layer is highly focused for high frequency spintronics due to a large emission power, narrow linewidth, and wide range tunability of the frequency \cite{kubota:13}.

In this paper, we investigate global synchronization in the one-dimensional (1-D) array $N$ serially connected and electrically coupled STNOs having perpendicular magnetic anisotropy in the presence of field-like torque by numerically solving the associated Landau-Lifshitz-Gilbert-Slonczewski(LLGS) equation. We prove that if the number of oscillators $N$ is low the field-like torque plays a crucial role in synchronizing them. On the other hand for large $N$, along with the field-like torque, the direction of the external field is also needed to be tuned slightly for the onset of complete synchronization.  The Kuramoto order parameter is computed to measure synchronization. The synchronization is discussed seperately for $N$ = 2, 10 and 100 STNOs. The tunability of the frequency of the synchronized oscillations for a wide range of the current and increment of the frequency by the field-like torque are also confirmed.   The stable synchronization due to the field-like torque is theoretically verified with the transverse Lyapunov exponent and it validates the numerical results.

\section{Model equation for 1-D array of $N$ serially connected STNOs}
We consider a system that consists of $N$ serially connected STNOs along with a load resistance $R_C$ as shown in Fig.\ref{fig1}.  Each oscillator consists of a perpendicularly magnetized free layer, where the direction of magnetization is allowed to change and an in-plane magnetized pinned layer where the direction of magnetization is fixed along the positive x-direction.   The unit vector along the direction of free layer's magnetization is given by ${\bf m}_j = (m_{jx},m_{jy},m_{jz})$, where $m_{jx}$, $m_{jy}$ and $m_{jz}$ are the $x$, $y$, and $z$ components of the unit magnetization vector ${\bf m}_j$, $j$ = 1,2,3,...,$N$, respectively.  The z axis is kept perpendicular to the plane of the free layer and ${\bf e}_{x}$,${\bf e}_{y}$ and ${\bf e}_{z}$ are unit vectors along the positive x,y and z directions, respectively. The unit vector along the direction of magnetization of the pinned layers is given by ${\bf p}(={\bf e}_x)$. The LLGS equation that governs the magnetization of the free layer of the $j$-th STNO is given by \cite{slon:96,arun_ieee,lakprl,lak}  
\begin{align}
&\frac{d{\bf m}_j}{dt}=-\gamma ~{\bf m}_j\times{\bf H}_{eff,j}+ \alpha ~{\bf m}_j\times\frac{d{\bf m}_j}{dt} \nonumber\\
&+\gamma \mu I_{j}~ {\bf m}_j\times ({\bf m}_j\times{\bf p})+\gamma \mu \beta I_{j}~ {\bf m}_j \times{\bf p},~j=1,2,...,N \label{goveqn}
\end{align}

 Here ${\bf H}_{eff,j}$ is the effective field, given by ${\bf H}_{eff,j} = H_a \sin\theta_h {\bf e}_x + [H_a \cos\theta_h + (H_k - 4\pi M_s) m_{zj}] {\bf e}_z$, which includes the externally applied magnetic field $H_a$, magnetocrystalline anisotropy field $H_k$ and shape anisotropy field (or demagnetizing field) $4\pi M_s$. The quantity $\theta_h$ is the angle of the externally applied field from ${\bf e}_z$, which may be denoted as the field-angle.  $M_s$ is the saturation magnetization, $\gamma$ is the gyromagnetic ratio, $\alpha$ is the Gilbert damping parameter, $\beta$ is the strength of the field-like torque and $\mu$ is  given by
\begin{equation}
\mu = \frac{\hbar\eta }{2 e M_s V(1+\lambda {\bf m}_j\cdot{\bf p})}.\label{mu}
\end{equation}
In Eq.\eqref{mu} $\hbar = h/2\pi$ ($h$ - Planck's constant), $V$ is the volume of the free layer, $e$ is the electron charge,  $\eta$ and $\lambda$ are dimensionless parameters which determine the magnitude and the angular dependence of the spin transfer torque, respectively. The current through the $j$-th STNO is given by \cite{tur1}
\begin{align}
I_{j}=I_{dc} [1-\sum_{i=1}^{N} \beta_{\Delta R_{i}} ({\bf m}_{i} \cdot {\bf p})]^{-1},\label{Ij}
\end{align}
where $I_{dc}$ and $\beta_{\Delta R_{i}}$ take the form 
 \begin{align}
 I_{dc}=\frac{ R_{c} I_{0}}{R_{c}+\sum_{i=1}^{N} R_{0i}} ~~~~{\rm and}~~~~~~~~~\beta_{\Delta R_{i}}=\frac{\Delta R_{i}} {R_{c}+\sum_{i=1}^{N} R_{0i}}. \nonumber
\end{align}
The term $\sum_{i=1}^{N} R_{0i}$ indicates the all-to-all coupling between the $N$ STNOs.  The input $I_0$ is a known DC current. Here, $R_{0i}=({R_{Pi}+R_{APi}})/{2}$ and $\Delta R_{i}=({R_{APi}-R_{Pi}})/{2}$. The quantities $R_{Pi}$ and  $R_{APi}$ are the resistances of the $i$-th STNO in the parallel and anti-parallel magnetization states, respectively.

To understand the dynamics of the magnetization of the free layer,  Eq.\eqref{goveqn} is numerically solved by Runge-Kutta 4th order method for the material parameters~\cite{tomo:18,kubota:13,tani:13} $M_s = 1448.3$ emu/c.c., $H_k = 18.6$ kOe, $\eta$ = 0.54, $\lambda$ = $\eta^2$, $\gamma$ = 17.64 Mrad/(Oe\hskip0.03cm s), $\alpha$ = 0.005, and $V=\pi \times 60 \times 60 \times 2$ nm$^3$.  Throughout our study $H_a$ is fixed as 2.0 kOe. Also, the values of $R_C$, $R_{0i}$ and $\Delta R_{i}$ are maintained as 50 $\Omega$, 0.1 $\Omega$ and 0.03 $\Omega$, respectively \cite{tur1}.

\section{Synchronization of 2 STNOs}
\begin{figure}[htp]
	\centering\includegraphics[angle=0,width=1\linewidth]{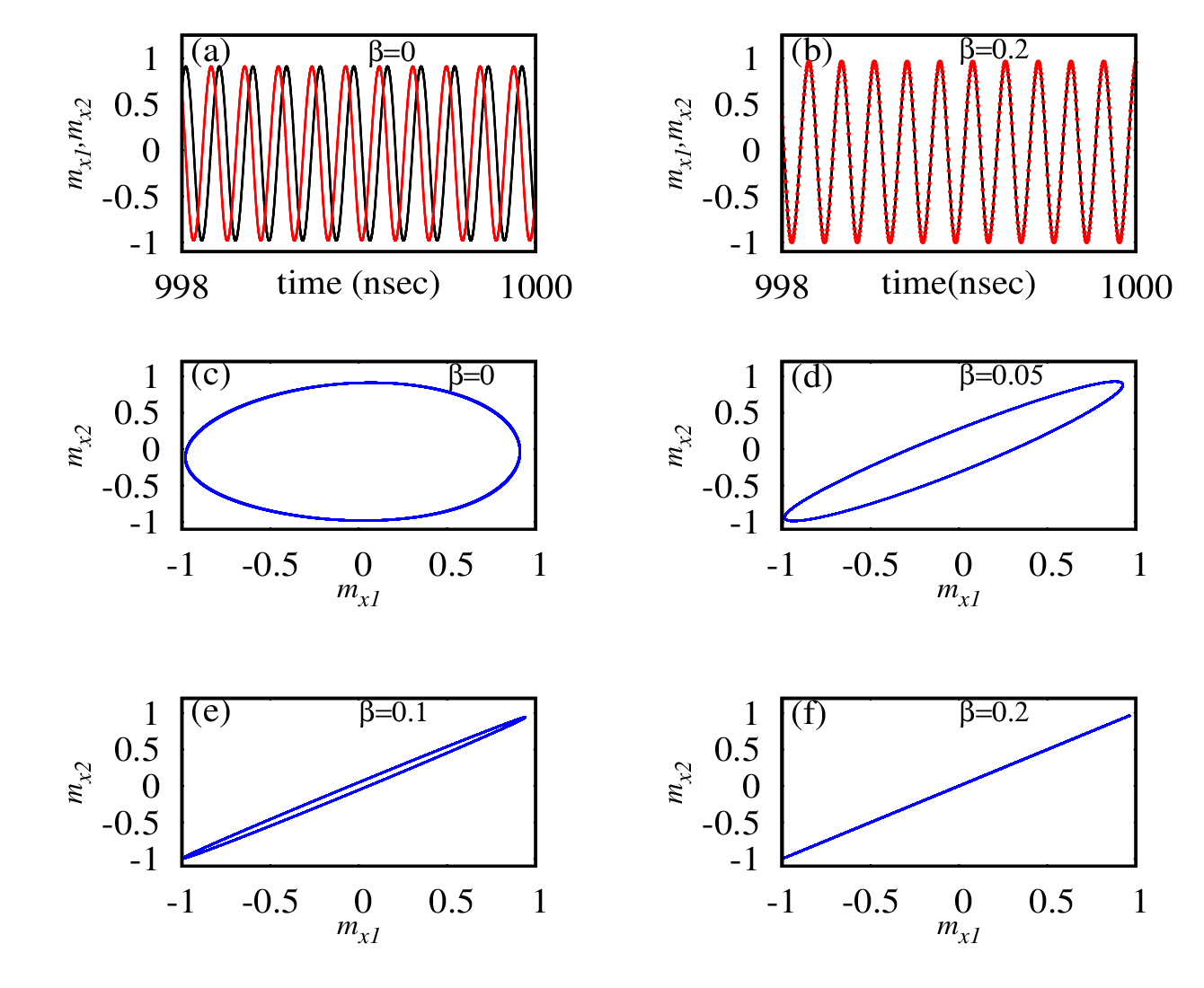}
	\caption{(a) Time evolutions of $m_{x1}$ (black solid line) and $m_{x2}$ (red solid line) for  $\beta$ = 0 and (b) time evolutions of $m_{x1}$ (black solid line) and $m_{x2}$ (red bullet points) for  $\beta$ = 0.2.  Phase portrait between $m_{x1}$ and $m_{x2}$ for (c) $\beta$ = 0, (d) $\beta$ = 0.05, (e) $\beta$ = 0.1 and (f) $\beta$ = 0.2.  Here $I_0$ = 2.5 mA, $\theta_h$ = 0$^\circ$ and $N$ = 2.}
	\label{fig2}
\end{figure}
To begin with, we consider two serially connected STNOs in the circuit along with a load resistance $R_C$ as shown in Fig.\ref{fig1}.  The corresponding LLGS equation for the two coupled STNOs is given by Eq.\eqref{goveqn} with $N$ = 2. To begin with the steady oscillations are obtained in the absence of the field-like torque by the current $I_0$ = 2.5 mA and $\theta_{h}$ = 0$^\circ$. The time evolutions of the x-component of the magnetization of the two oscillators $m_{x1}$ (black solid line) and $m_{x2}$ (red solid line) are plotted in Fig.\ref{fig2}(a) for the duration between the times 998 ns and 1000 ns. It is observed that the oscillations of the two STNOs are not synchronized with each other.  However, in the presence of the field-like torque the oscillators get synchronized and exhibit synchronized oscillations.  The synchronized oscillations of $m_{x1}$ (black solid line) and $m_{x2}$ (red bullets) are confirmed in Fig.\ref{fig2}(b), where the strength of the field-like torque $\beta$ is 0.2 while $I_0$ = 2.5 mA and $\theta_h$ = 0$^\circ$. The initial conditions for the two STNOs are taken near positive $z$-axis since the anisotropy is perpendicular along the $z$-axis.  From Figs.\ref{fig2}(a) and (b) we can understand that the system of 2 STNOs can be synchronized by the field-like torque. 

To observe the impact of increase in the strength of the field-like torque on synchronization, the phase portrait plots are plotted between the $x$-components of the unit magnetization vectors ${\bf m}_1$ and ${\bf m}_2$, namely $m_{x1}$ and $m_{x2}$, in Figs.\ref{fig2}(c-f)  for  different values of $\beta$ = 0, 0.05, 0.1 and 0.2, respectively.  When the value of $\beta$ is 0, the curves in the $m_{x1}$ and $m_{x2}$ plane are more separated  as shown  in Fig.\ref{fig2}(c). The separation is reduced as the strength of field-like torque is increased, see Figs.\ref{fig2}(d-e), and when it is 0.2 there is no separation between $m_{x1}$ and $m_{x2}$ as shown in Fig.\ref{fig2}(f). The straight line at 45$^\circ$ with respect to $m_{x1}$ means that the $m_{x1}$ and $m_{x2}$ are perfectly synchronized with each other.  Figs.\ref{fig2}(c-f) confirm that the field-like torque can bring the coupled oscillators towards their synchronization. 
\begin{figure}[htp]
	\centering\includegraphics[angle=0,width=1.0\linewidth]{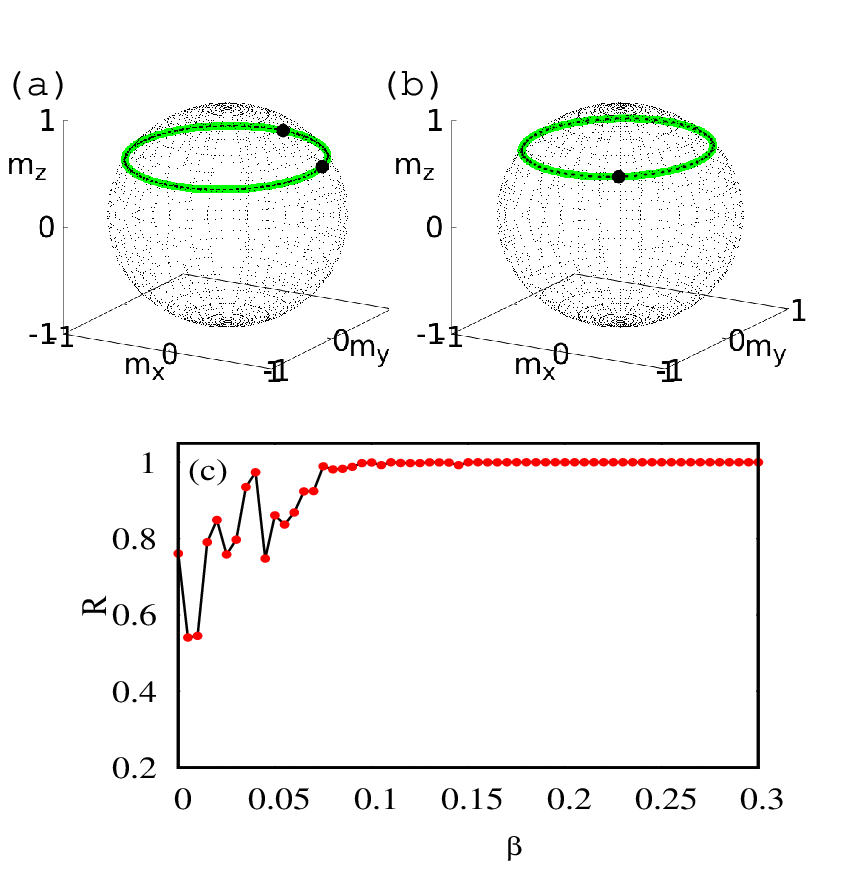}
	\caption{Magnetization trajectories (between $t$ = 999 ns and $t$ = 1000 ns) of ${\bf m}_1$ (green solid line) and ${\bf m}_2$ (black dotted line) with their corresponding instantaneous magnetization states (black bullets) at $t$ = 1000 ns for (a) $\beta$ = 0 and (b) $\beta$ = 0.2. (c) Order parameter $R$ against the strength of field-like torque $\beta$.  Here $I_0$ = 2.5 mA, $\theta_h$ = 0$^\circ$ and $N$ = 2.}
	\label{fig3}
\end{figure}
From the Figs.\ref{fig2}(a) we can note that the range and frequency are the same for both of the oscillators. It means that there is a lag between the two oscillators in the absence of field-like torque.  It can also be observed in Figs.\ref{fig3}(a) and (b) where the trajectories of the magnetizations  of the two STNOs are plotted in green solid line for $\beta$ = 0 and $\beta$ = 0.2, respectively. The black dotted lines plotted in the green lines of Figs.\ref{fig3}(a) and (b) are the respective trajectories of the second oscillator. The two black colour bullet points in the trajectories corresponding to $\beta$ = 0 indicate the instantaneous magnetization states of the two oscillators at time $t$ = 1000 ns. It clearly shows that the two oscillators follow the same trajectory in the absence of the field-like torque, whereas for $\beta$ = 0.2 the magnetization of the two oscillators take exactly the same position at time $t$ = 1000 ns. Fig.\ref{fig3}(a) confirms that the presence of field-like torque can synchronize the desynchronized oscillations of the STNOs.
The synchronization between the STNOs can also be identified by the Kuramoto order parameter, also called the coherence parameter which is given by
\begin{align}
R = \frac{1}{N} |\sum_{k=1}^{N} e^{i\phi_{k}}|, ~~~0<R<1, \notag
\end{align}
where $\phi_{k} = \tan^{-1}(m_{yk}/m_{xk})$ is the azimuthal angle of the magnetization of the $k$-th STNO in the plane of ${\bf e}_x$ and ${\bf e}_y$.
Direct calculations can show that when $R$ = 0 the STNOs are asynchronous and when $R$ = 1 they exhibit complete synchronization. To visualize the impact of the field-like torque on synchronization the order parameter $R$ is plotted against the  strength of field-like torque $\beta$ when $I_0$ = 2.5 mA and $\theta_h$ = 0$^\circ$ in Fig.\ref{fig3}(b), where we can clearly see that the value of $R$ is well below 1.0 at $\beta$ = 0 and increases with $\beta$ and reaches the value 1.0 around $\beta$ = 0.1. This means that in the absence of field-like torque, the oscillations are asynchronous and can be synchronized by suitable strength of the field-like torque. As shown in Fig.\ref{fig3}(b) that there is no gradual increase in $R$ with $\beta$ towards 1.0, which is due to the fact that the initial conditions for every value of $\beta$ have been taken randomly around ${\bf e}_z$ without maintaining any uniformity in their choice.

\begin{figure}[htp]
	\centering\includegraphics[angle=0,width=1.0\linewidth]{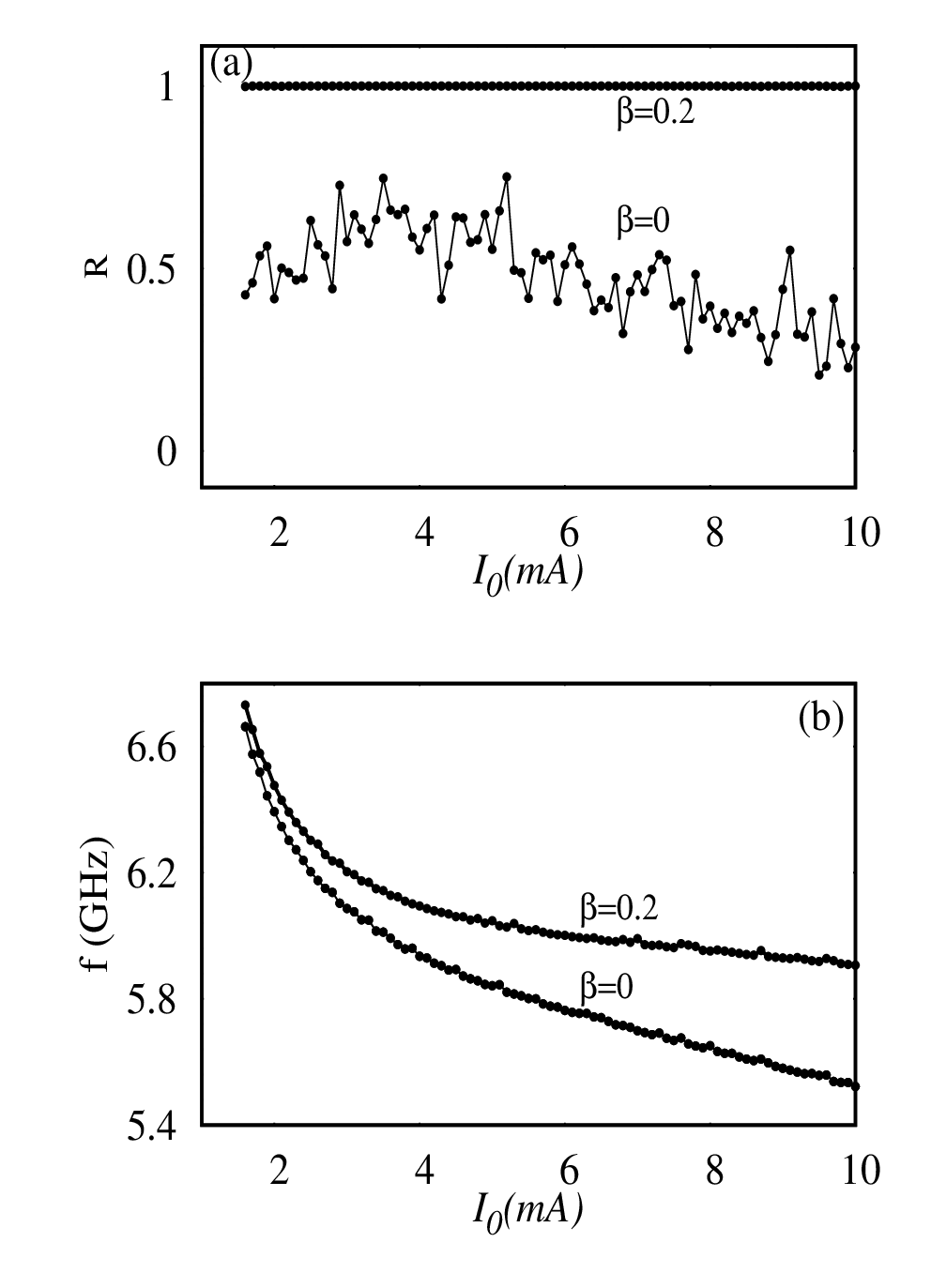}
	\caption{(a) The order parameter $R$ and (b) the frequency $f$ against current $I_0$ for 2 STNOs  when $\beta$ = 0 and $\beta$ = 0.2. Here $\theta_h$ = 0$^\circ$ and $N$ = 2.}
	\label{fig4}
\end{figure}

To confirm the onset of synchronization and the essentiality of the field-like torque for synchronization over a wide range of currents, in Fig.\ref{fig4}(a), the order parameter $R$ is plotted against the current for $\beta$ = 0 and 0.2 when $I~>$ 1.6 mA and $\theta_h$ = 0$^\circ$ as no oscillations occur below $I_0$ = 1.6 mA. It is observed that  the two STNOs exhibit asynchronous oscillations in the absence of field-like torque for all the currents. This can be confirmed by the fact that the values of $R$ are well below 1.0 for $\beta$ = 0. Also, the tunability of the current alone, without the field-like torque, does not lead to synchronization between the two oscillators.  When $\beta$ = 0.2, the values of $R$ for $I_0 > 1.6$ mA equals 1.0, which implies that the two STNOs can exhibit synchronous oscillations due to the field-like torque. Figures \ref{fig3}(b) and \ref{fig4}(a) confirm that complete synchrony is achieved by the field-like torque.
The frequency of the two STNOs are plotted when the two STNOs are (i) desynchronized ($\beta$ = 0) and (ii) synchronized ($\beta$ = 0.2) in Fig.\ref{fig4}(b). As observed in Fig.\ref{fig2}(a) the frequencies of the two oscillators in the absence of field-like torque are the same.  From Fig.\ref{fig4}(b) we can observe that the frequencies of the two STNOs are increased due to the field-like torque. This implies that in addition to achieving complete synchronization the field-like torque also enhances the frequencies of the synchronized STNOs.

\section{Synchronization of 10 STNOs}
\begin{figure}[htp]
	\centering\includegraphics[angle=0,width=1\linewidth]{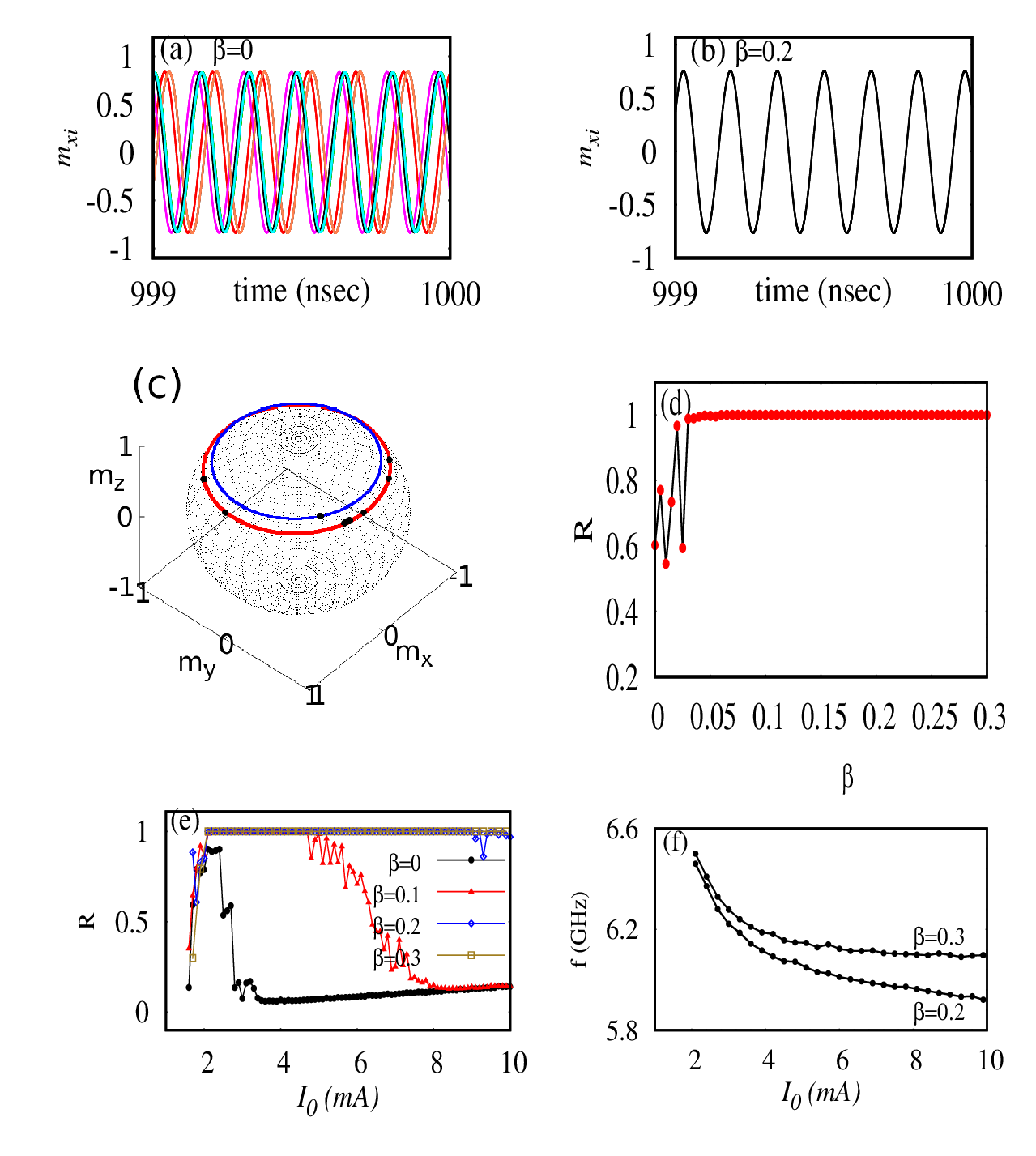}
	\caption{Time evolutions of $m_{xi}$, $i$ = 1,2,...,10 for (a) $\beta$ = 0 and (b) $\beta$ = 0.2. (c) Magnetization trajectories (between $t$ = 999 ns and $t$ = 1000 ns) of ${\bf m}_i$, $i$ = 1,2,...,10, with their corresponding instantaneous magnetization states (black bullets) at $t$ = 1000 ns for $\beta$ = 0 (red solid line) and $\beta$ = 0.2 (blue solid line).  Variation of order parameter $R$ against (d) the coefficient of field-like torque for $I_0$ = 2.5 mA and (e) the current. (f) Frequency with respect to the current.  Here $\theta_h$ = 0$^\circ$ and $N$ = 10. }
	\label{fig5}
\end{figure}
To check whether the field-like torque is able to synchronize  when more STNOs are added within the system, the number of STNOs, $N$, is increased from 2 to 10.  The corresponding LLGS equation is again given by Eq.\eqref{goveqn} but now with $N$ = 10.  The time evolutions of $m_{x1},m_{x2},...,m_{x10}$ are plotted in Fig.\ref{fig5}(a) and (b) in the absence and presence of the field-like torque, respectively, when $I_0$ = 2.5 mA and $\theta_h$ = 0$^\circ$. From Fig.\ref{fig5}(a) we can understand that in the absence of the field-like torque, there exists no complete synchronization  between them.  However, when the field-like torque is present the asynchronized oscillations get synchronized as confirmed in Fig.\ref{fig5}(b). The magnetization trajectories of the ten oscillators are plotted in Fig.\ref{fig5}(c) for $\beta$ = 0 as red solid line and for $\beta$ = 0.2 as blue solid line.  The black bullets are the instantaneous magnetization states of the ten oscillators at $t$ = 1000 ns on their respective trajectories.  The ten oscillators follow the same trajectory in the absence of the field-like torque. If we count the number of black bullets on the trajectory corresponding to $\beta$ = 0 we will end up with 7 instead of 10.  It implies that among the ten oscillators, two or more oscillators can synchronize individually and form clusters within the system. However, when the field-like torque is applied all the oscillators are completely synchronized as shown by the single black bullet on the trajectory corresponding to $\beta$ = 0.2. The order parameter is plotted in Fig.\ref{fig5}(d) against $\beta$ when $I$ = 2.5 mA and $\theta_h$ = 0$^\circ$.  It is clearly visible that the value of the $R$ reaches the value 1.0 when $\beta\geq$ 0.05, which means that a small strength of the field-like torque can completely synchronize the ten oscillators. Again the zig-zag increment of $R$ against $\beta$ before reaching the value 1.0 is due to the randomness of the choice of initial conditions for the ten oscillators at each vale of $\beta$. 

To confirm the synchronization of the 10 STNOs for a wide range of current due to the field-like torque, the order parameter is plotted against the current for different values of $\beta$ = 0, 0.1, 0.2 and 0.3 in Fig.\ref{fig5}(e) while $\theta_h$ = 0$^\circ$.  From this figure, we can realize that complete synchronization is not possible by increasing the current alone without the field-like torque.  If we increase the strength of the field-like torque from $\beta$ = 0 to 0.3, the range of the current at which the synchronization is possible increases. Also, we can observe that the synchrony is lost at higher values of the current for lower strengths of the field-like torque. However when $\beta$ = 0.3 we can see that the complete synchronization is achieved even at higher values of current. The frequency for the ten STNOs at their synchronization state is plotted in Fig.\ref{fig5}(f) for two different values of $\beta$ = 0.2 and 0.3 and it confirms that the frequency of the synchronized oscillations can be increased by the strength of the field-like torque.
\section{Synchronization of 100 STNOs}
\begin{figure}[htp]
	\centering\includegraphics[angle=0,width=1\linewidth]{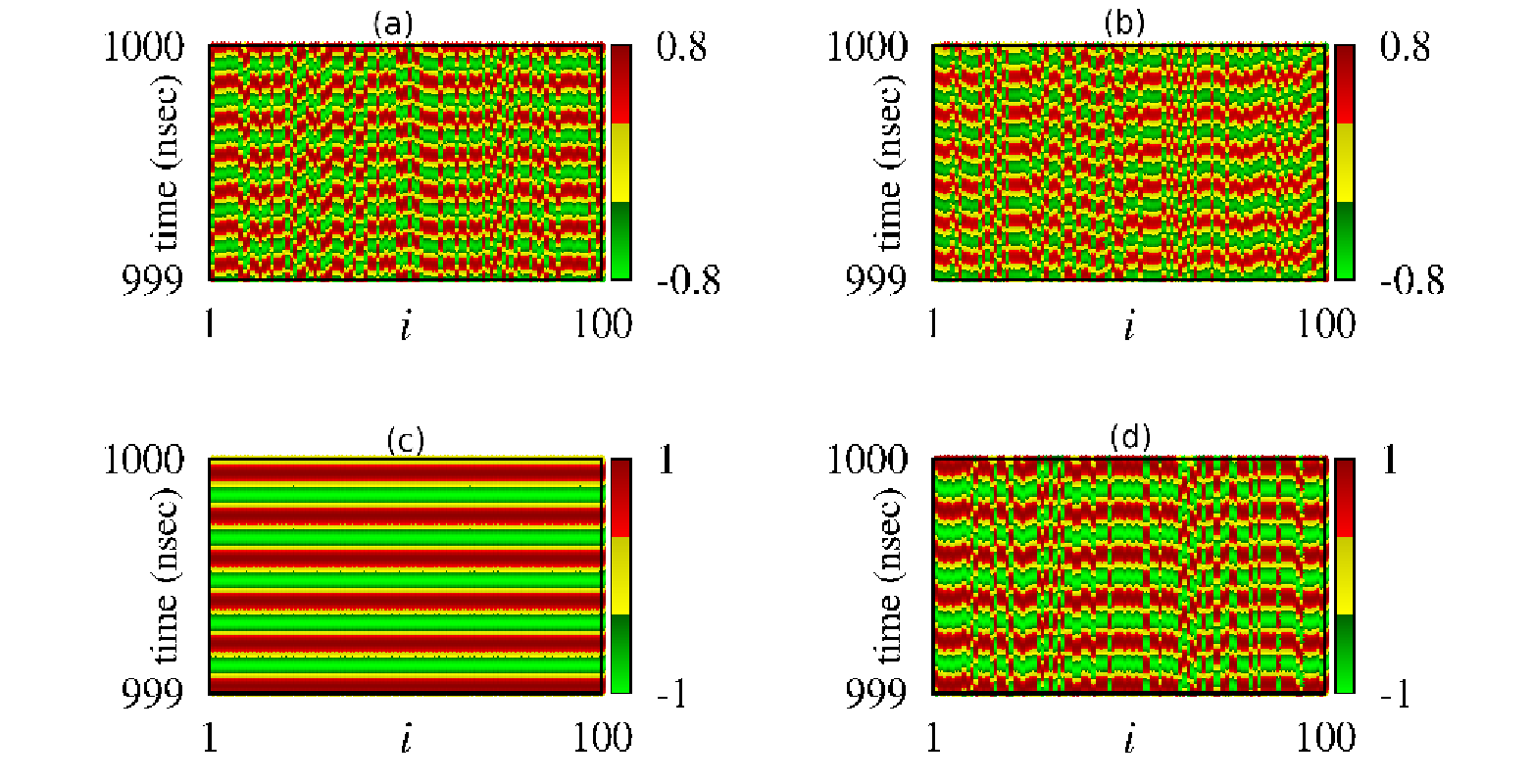}
	\caption{Spatio-temporal for $m_{xi}$, $i$ = 1,2,...,100 for (a) $\beta$ = 0.0 and $\theta_h$ = $0^\circ$, (b) $\beta$ = 0.2 and $\theta_h$ = $0^\circ$, (c) $\beta$ = 0.2 and $\theta_h$ = $6^\circ$ and (d) $\beta$ = 0 and $\theta_h$ = $6^\circ$. Here $I_0$ = 2.5 mA and $N$ = 100.}
	\label{fig6}
\end{figure}
To validate the synchronization due to the field-like torque in a large 1-D array of STNOs, the system is now enhanced to $N$ = 100.  The governing dynamical equation is then again given by Eq.\eqref{goveqn} but with $N$ = 100. The spatio-temporal plots for the $m_x$ components of all the 100 oscillators are plotted in Figs.\ref{fig6}(a) and (b) for $\beta$ = 0 and 0.2, respectively when $I_0$ = 2.5 mA and $\theta_h$ = 0$^\circ$.  It is clearly identified that complete synchronization is not achieved due to the field-like torque even with the strength $\beta$ = 0.2. We have verified that there is no synchronization even if the strength of the field-like torque is increased above 0.2 (results not shown here).  However, the presence of the field-like torque  leads to the formation of different clusters (i.e. different groups are formed within the system and complete synchronization is exhibited within some of these individual groups) within the system as shown in Fig.\ref{fig6}(b).  However, when the direction of the external magnetic field is slightly tilted from $\theta_h$ = 0$^\circ$ to $\theta_h$ = 6$^\circ$, in the presence of field-like torque strength $\beta$ = 0.2, all the oscillators get completely synchronized with each other as shown in Fig.\ref{fig6}(c). Unlike the previous cases where the synchronization was achieved only with the field-like torque  alone (and $\theta_h$ = 0) for N = 2 and 10, the system of 100 STNOs now requires a slight tunability  in the direction of the external magnetic field in addition to the presence of field-like torque.  One may doubt that the tunability in the direction of the magnetic field alone can bring complete synchronization without the introduction of field-like torque.  To verify  this possibility, the spatio-temporal plot for $I_0$ = 2.5 mA and $\theta_h$ = 6$^\circ$ is plotted in the absence of the field-like torque.  It can be clearly observed from Fig.\ref{fig6}(d) that the tunability in the direction of the external field only succeeds to form clusters within the system. Also, it has been verified that there is no complete synchronization in the entire range  $\theta_h$ = 0$^\circ$ - 180$^\circ$ in the absence of the field-like torque (results not shown here). Figs.\ref{fig6}(a-d) clearly demonstrate the essentiality of the field-like torque as well as the tunability in the direction of the external field for achieving complete synchronization of 100 STNOs.  
\begin{figure}[htp]
	\centering\includegraphics[angle=0,width=1\linewidth]{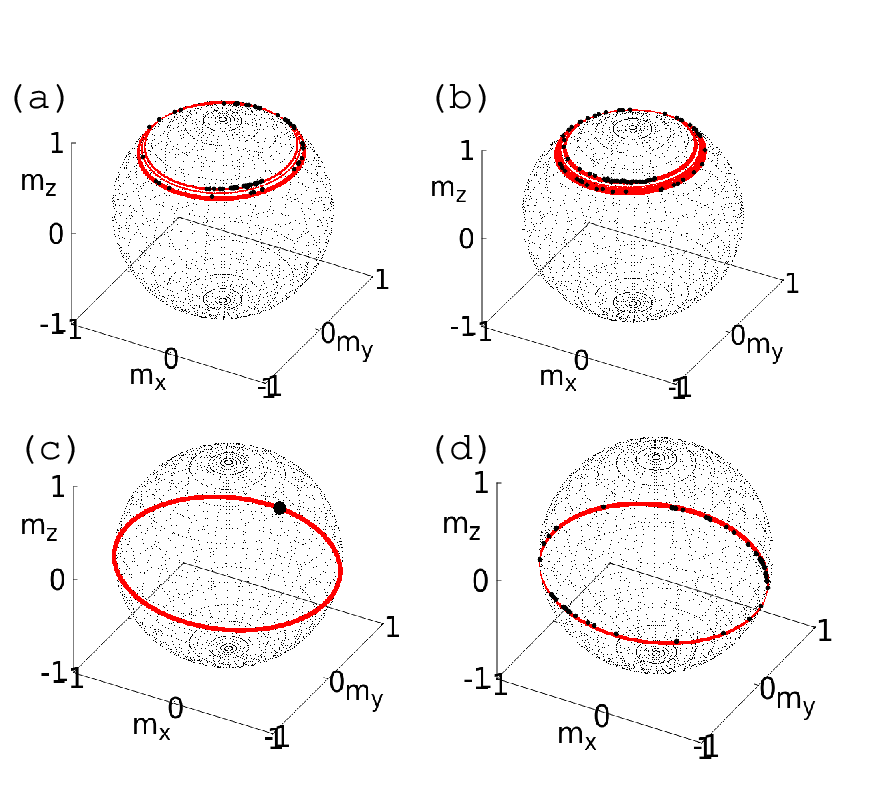}
\caption{ Magnetization trajectories (between $t$ = 999 ns and $t$ = 1000 ns) of ${\bf m}_i$, $i$ = 1,2,...,100, with their corresponding instantaneous magnetization states (black bullets) at $t$ = 1000 ns for (a) $\beta$ = 0.0, $\theta_h$ = $0^\circ$, (b) $\beta$ = 0.2, $\theta_h$ = $0^\circ$, (c) $\beta$ = 0.2, $\theta_h$ = $6^\circ$ and (d) $\beta$ = 0, $\theta_h$ = $6^\circ$. Here $I_0$ = 2.5 mA and $N$ = 100.}
	\label{fig7}
\end{figure}

\begin{figure*}[htp]
	\centering\includegraphics[angle=0,width=0.8\linewidth]{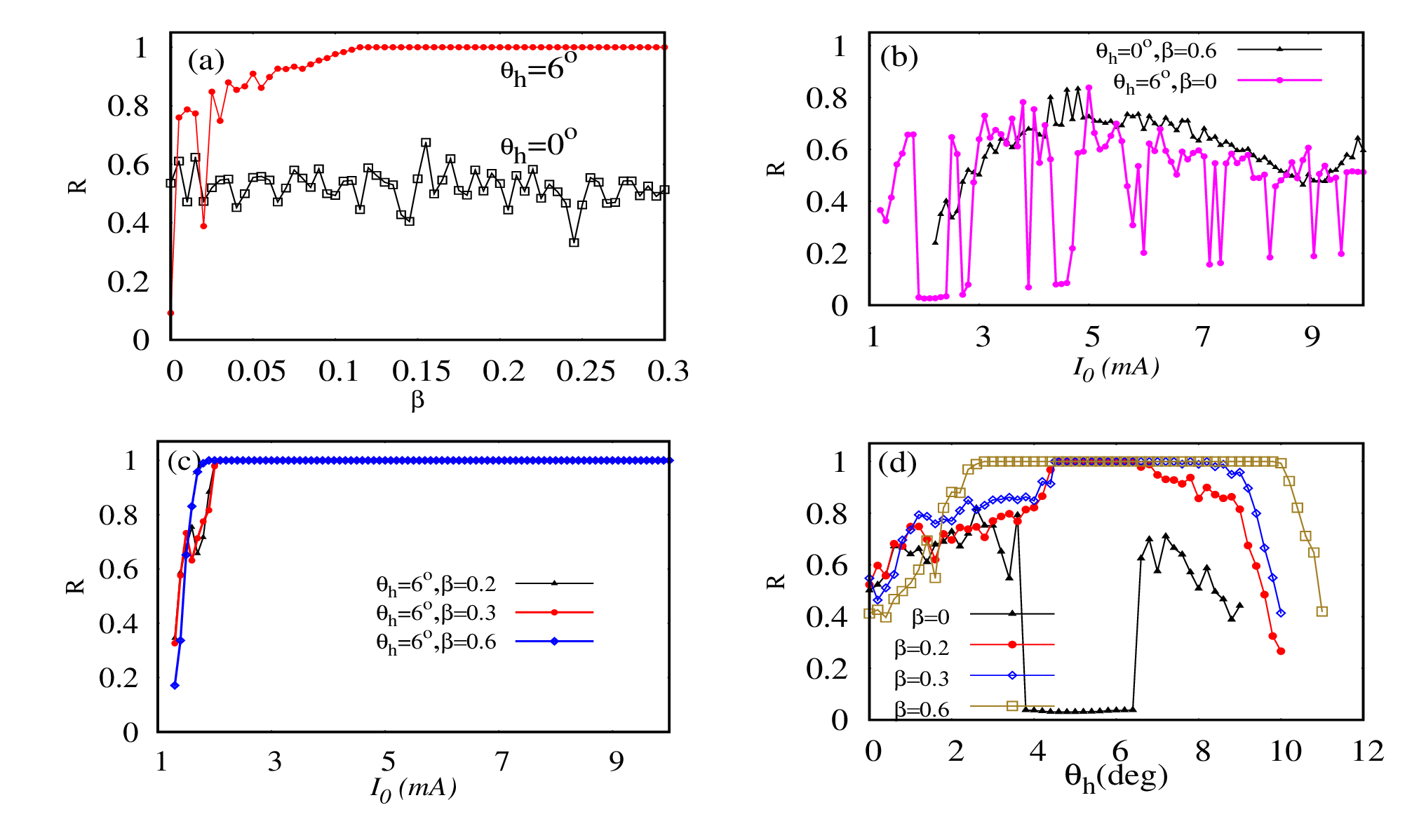}
	\caption{The order parameter $R$ for 100 STNOs against (a) $\beta$ for $\theta_h$ = 0$^\circ$ and 6$^\circ$ while $I_0$ = 2.5 mA, { and against current for (b) ($\theta_h,\beta$) = (0$^\circ$,0.6), (6$^\circ$,0) and (c) ($\theta_h,\beta$) = (6$^\circ$,0.2), (6$^\circ$,0.3) and (6$^\circ$,0.6), and against (d) field-angle $\theta_h$ for $\beta$ = 0 (black, $\blacktriangle$), 0.2 (red, $\bullet$), 0.3 (blue, $\diamond$) and 0.6 (gold, $\square$) while $I_0$ = 2.5 mA. }}
	\label{fig8}
\end{figure*}
The magnetization trajectories corresponding to Figs.\ref{fig6}(a-d) are plotted in Figs.\ref{fig7}(a-d), respectively.  The black colour bullets on the trajectories indicate the magnetization state of each of the 100 oscillators at the time $t$ = 1000 ns.  The continuous presence of the black bullets on at the trajectory of Fig.\ref{fig7}(b) indicates that the field-like torque forms a group of closely oscillating STNOs.  Also, the total number of the black butllets in the trajectory is less than 100, which implies the formation of clusters within the system.  The single black  bullet in the trajectory in Fig.\ref{fig7}(c)  confirms the complete synchronization of the 100 STNOs due to the change in the direction of the  external field from 0$^\circ$ to 6$^\circ$ and the presence of the field-like torque with $\beta$ = 0.2. It also confirms that the same magnetization state is acquired by all the 100 STNOs.  Apart from achieving complete synchronization, the resultant synchronized oscillations exhibit a large amplitude of the oscillations as shown in Fig.\ref{fig7}(c). Figs.\ref{fig6}(d) and \ref{fig7}(d) plotted for $N$ = 100, $\beta$ = 0, $I_0$ = 2.5 mA and $\theta_h$ = 6$^\circ$ affirm that {only clusters are formed within the system by varying the field-angle in the absence of the field-like torque.}

To visualize the impact of the field-like torque as well as the field-angle $\theta_h$ on the synchronization of the serially coupled STNOs, Figs.\ref{fig8}(a-d) are plotted for the order parameter $R$.  In Fig.\ref{fig8}(a) the order parameter is plotted against the strength of the field-like torque for $\theta_h$ = 0$^\circ$ and 6$^\circ$ when $I_0$ = 2.5 mA. When the field-angle is 0$^\circ$ the order parameter is well below 1 for the strengths of the field-like torque $\beta$ in the entire range from 0 to 0.3. On the other hand, when the field-angle is fixed as 6$^\circ$ the order parameter reaches the value 1 for $\beta\geq$0.11. Thus, one may note that the synchronization by the field-like torque is not possible with the external field applied along the $z$-axis (so that the $\theta_h$ = 0$^\circ$). However, a slight tunability in the field-angle to $\theta_h = 6^\circ$ induces a complete synchronization between the oscillators. {Figure \ref{fig8}(b) shows the variation of the order parameter against the current, and one may note that the value of the order parameter does not reach the value of 1 even for a large strength of the field-like torque, $\beta$ = 0.6, when $\theta_h$ = 0$^\circ$ and for the field-angle $\theta_h$ = 6$^\circ$ when $\beta$ = 0.}  This figure confirms that complete synchronization can be achieved only when tuning the field-angle in the presence of the field-like torque. 
\begin{figure}[htp]
	\centering\includegraphics[angle=0,width=0.8\linewidth]{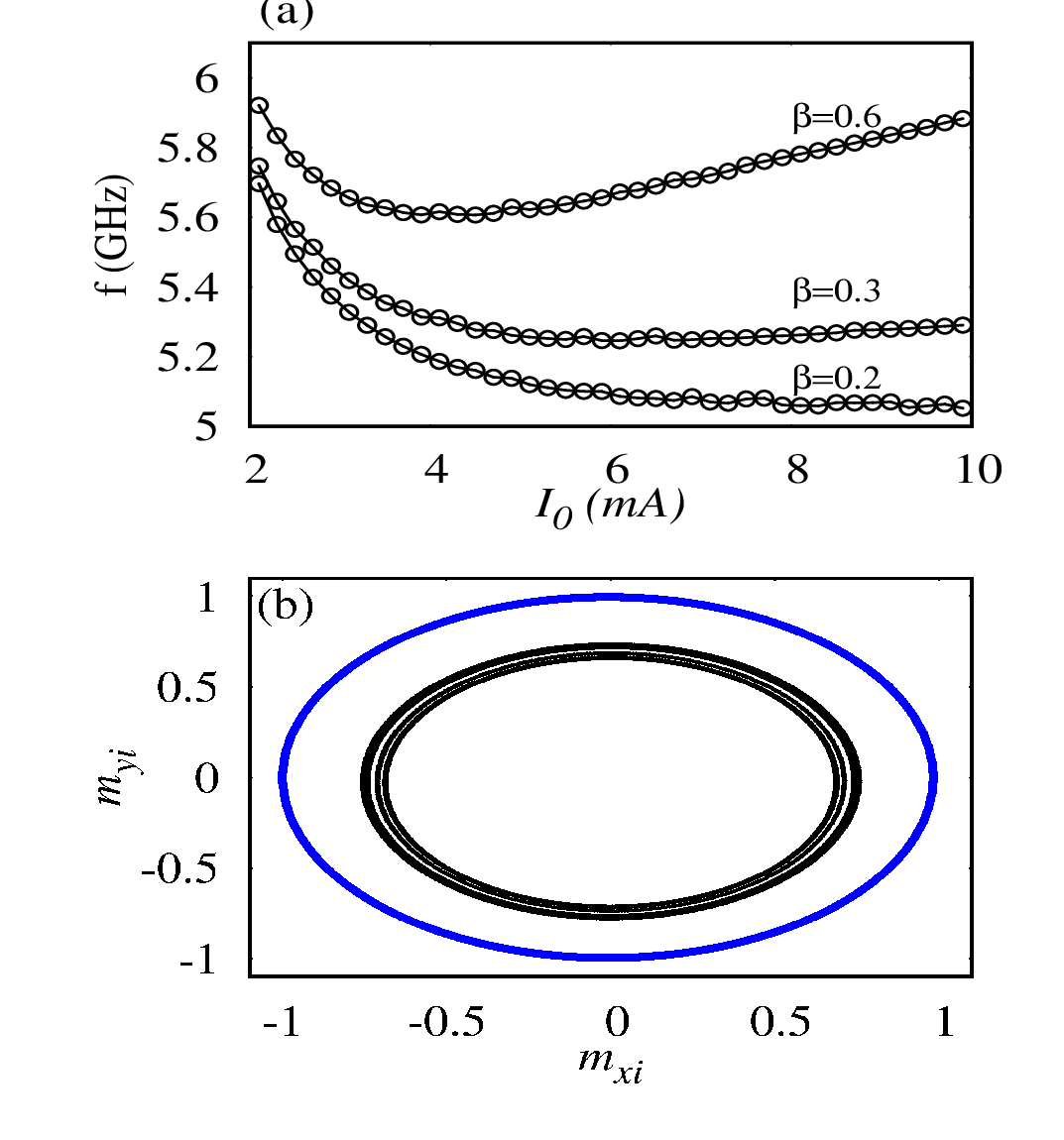}
	\caption{(a) Frequency against the current for $\beta$ = 0.2, 0.3 and 0.6 for $N$ = 100. (b) phase portrait of 100 STNOs between $m_{xi}$ and $m_{yj}$ for $\beta = 0,~ \theta_h = 0^\circ$ (black) in the cluster state and for $\beta = 0.2,~ \theta_h = 6^\circ$ (blue) in the synchronized state when $I_0$ = 2.5 mA.}
	\label{fig9}
\end{figure}

To verify the existence of synchronization over a wide range of current, the order parameter is computed for currents ranging from 1.2 mA to 10.0 mA and it is plotted in Fig.\ref{fig8}(c) for different values of $\beta$ = 0.2, 0.3 and 0.6 while $\theta_h$ = 6$^\circ$. We can observe from this figure that the oscillators are not synchronized when the value of current is low and they get synchronized for $I\geq$2.1 mA when $\beta$ = 0.2 and 0.3 and for $I\geq$1.9 mA when $\beta$ = 0.6. It evidences that the increase of the field-like torque reduces the magnitude of the current for the onset of the synchronized oscillations. Figure \ref{fig8}(c) confirms the  synchronization between the 100 STNOs  for a wide range of currents by tilting the external field in the presence of field-like torque.

Further, to investigate the range of the field-angle $\theta_h$ for which synchronization is possible, Fig.\ref{fig8}(d) is plotted  between the order parameter $R$ and the angle $\theta_h$ for different values of $\beta$ = 0, 0.2, 0.3 and 0.6 while $I_0$ = 2.5 mA.  From  Fig.\ref{fig8}(d) it can be observed that in the absence of the field-like torque, there is no possibility of synchronization by tuning the direction of the external field. If the strength of the field-like torque is increased to $\beta$ = 0.2 from 0, the oscillators are synchronized for the field-angles in the range 4.6$^\circ$ - 6.4$^\circ$.  When $\beta$ is increased further to 0.3 the range of field-angle includes 4.5$^\circ$ - 7.4$^\circ$ for complete synchronization.  The range of the field angle for the complete synchronization can be extended further to 2.8$^\circ$ - 10$^\circ$ by increasing the strength of the field-like torque to 0.6 as shown in Fig.\ref{fig8}(d). Hence, the above study indicates that  complete synchronization is not possible by considering the tunability of field-angle in the absence of field-like torque.  However, the presence of field-like torque induces the synchronized oscillations while the field-angle is tuned and it also enhances the range of field-angle for complete synchronization.

The frequencies of the 100 synchronized oscillators are plotted in Fig.\ref{fig9}(a) with respect to the current for different values of the strengths of field-like torque, $\beta$ = 0.2, 0.3 and 0.6.  The figure confirms the tunability in frequency for a wide range of currents. Also, an increase in the strength of the field-like torque enhances the frequency at a given current. 
%Moreover, it can be observed that at larger values of the current, the field-like torque increases the rate of change of frequency with respect to the current.  
Hence, in addition to synchronization the presence of the field-like torque also enhances the frequency of the synchronized oscillations.   The phase portraits between $m_{xi}$ and $m_{yi}$ for $I_0$ = 2.5 mA are plotted for (i) $\beta = 0,~ \theta_h = 0^\circ$ with black lines (cluster) and (ii) $\beta = 0.2,~ \theta_h = 6^\circ$ with blue line (synchronized state) in Fig.\ref{fig9}(b). It evidences that the large amplitude synchronized oscillations are exhibited for $m_x$ and $m_y$.   The above mentioned large amplitude of the synchronized oscillations, especially for larger number of coupled STNOs, may benefit towards achieving larger magnetoresistance and consequently high output power for the voltage oscillations \cite{zahe,tare}. We have verifed that the synchronization can also be achieved for more than 100 STNOs with the same electrical coupling discussed here (results not shown here). 
\begin{figure}
	\centering
	\includegraphics[width=0.7\linewidth]{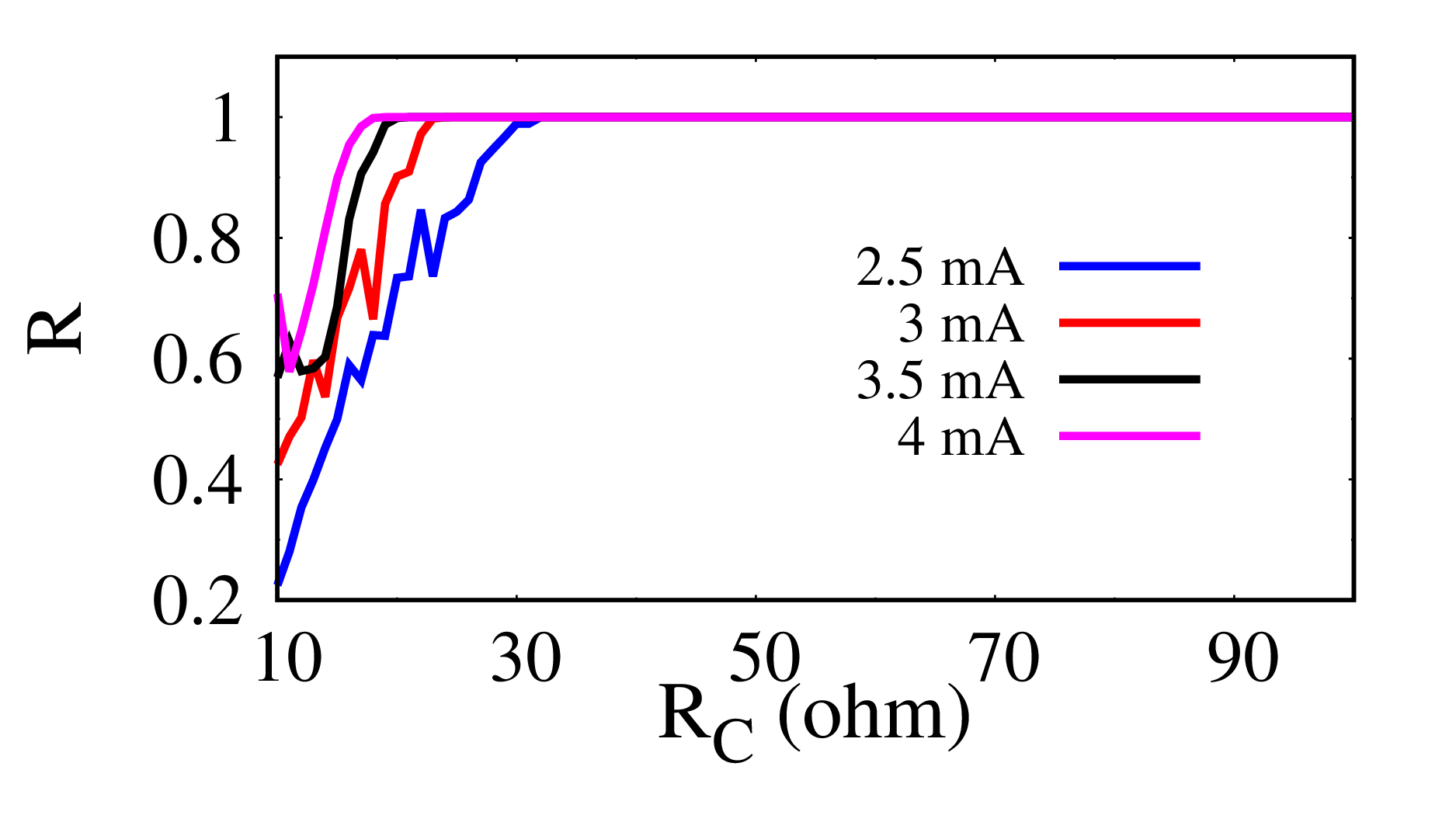}
	\caption{Variation of the order parameter $R$ against the load resistance $R_C$ for the different currents $I_0$ = 2.5, 3, 3.5 and 4 mA. Here, $N$ = 100, $\theta_h$ = 5$^\circ$ and $\beta$ = 0.6.}
	\label{Rc}
\end{figure}

 Finally, a question arises as to how the load resistance $R_C$ in Fig.\ref{fig1} affects the synchronization. To answer this question we have plotted the order parameter $R$ for $N$ = 100 against $R_C$ for different values of current $I_0$ =  2.5, 3, 3.5 and 4 mA when $\theta_h$ = 5$^\circ$ and $\beta$ = 0.6 in Fig.\ref{Rc}.  From the figure it can be observed that the value of the order parameter $R$ becomes 1 when the load resistance $R_C$ is above a critical value which is 30$\Omega$ here. This implies that the synchronization between STNOs can be maintained for a large range of load resistance as shown in the figure. Also, when the current is increased the minimum load resistance below which the synchronization is lost decreases below the critical value.
\section{Route to Synchronization}
\begin{figure}[htp]
	\centering\includegraphics[angle=0,width=1\linewidth]{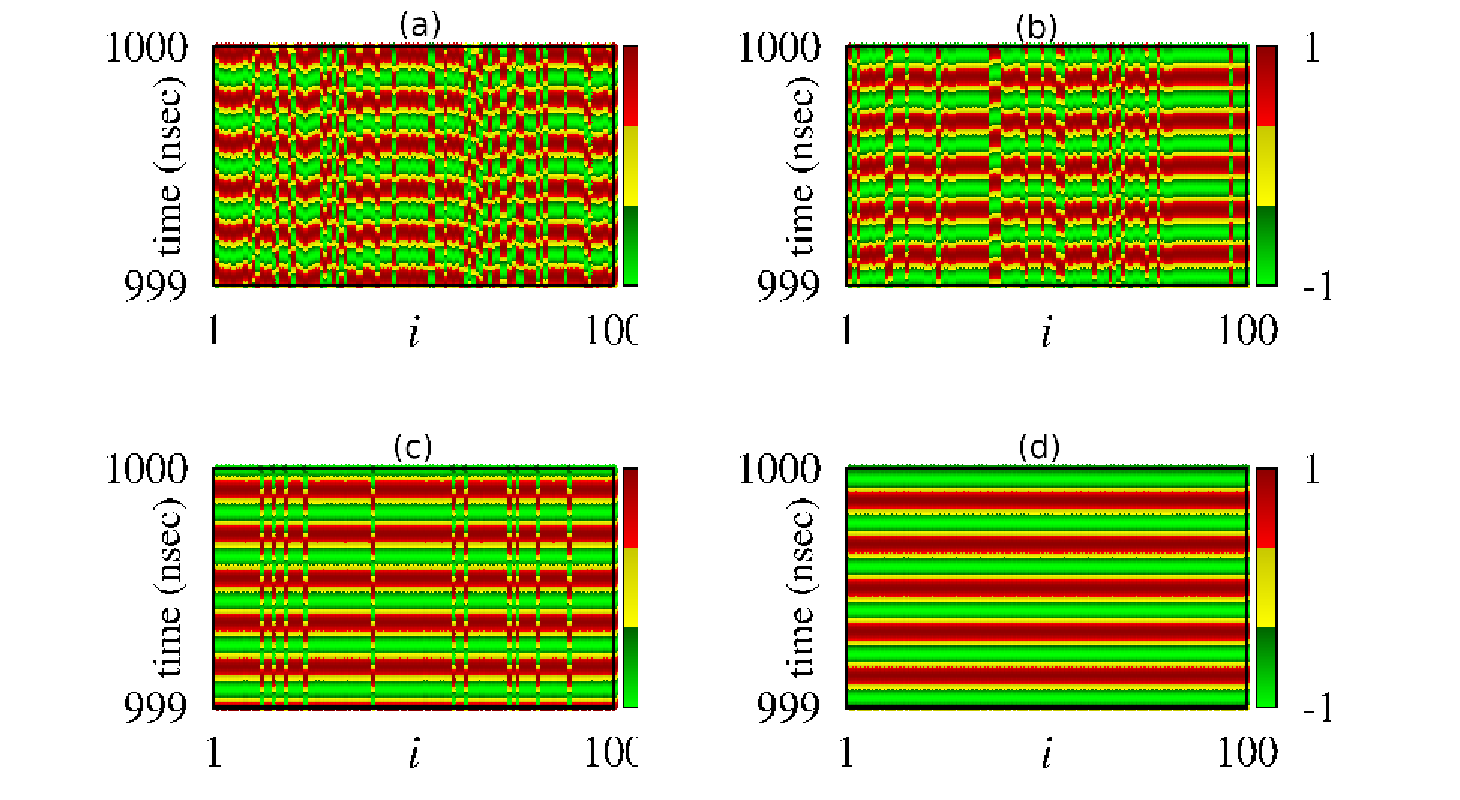}
	\caption{Spatio-temporal for $m_{xi}$, $i$ = 1,2,...,100 for (a) $\beta$ = 0.0, (b) $\beta$ = 0.03, (c) $\beta$ = 0.05 and (d) $\beta$ = 0.15. Here $I_0$ = 2.5 mA, $\theta_h=6^\circ$ and $N$ = 100.}
	\label{fig10}
\end{figure}
To discuss the way to reach the complete synchronization of the 100 coupled STNOs by the impact of field-like torque the spatio-temporal plots for the $m_{xi}$, $i=1,2,...,100$ are plotted for $\beta$ = 0, 0.03, 0.05 and 0.15  in Figs.\ref{fig10}(a-d), respectively, between the times 999 ns and 1000 ns while $I_0$ = 2.5 mA and $\theta_h$ = 6$^\circ$.  In Fig.\ref{fig10}(a) we can see that the oscillators are not synchronized without the field-like torque.  When the $\beta$ is increased to 0.03, clusters are formed within the system as shown in Fig.\ref{fig10}(b).  The number of clusters is reduced when the strength of the field-like torque is increased further to $\beta$ = 0.05.  In Fig.\ref{fig10}(c) we can identify that the number of clusters is reduced to 2 when compared to Fig.\ref{fig10}(b). This has indeed been checked with the phase trajectories of the spins where one can identify just two black bullets (which is not shown here).  When the value of $\beta$ is increased further to 0.15, the 100 STNOs are completely synchronized with each other as confirmed in Fig.\ref{fig10}(d).  Hence the Figs.\ref{fig10}(a-d) confirm the fact that the presence of the field-like torque induces  clusters within the system and the increment in the strength of $\beta$ reduces the number of clusters and finally complete synchronization is achieved when the strength of the filed-like torque crosses a threshold value.

\section{Impact of field-angle}
\begin{figure}[htp]
	\centering\includegraphics[angle=0,width=1\linewidth]{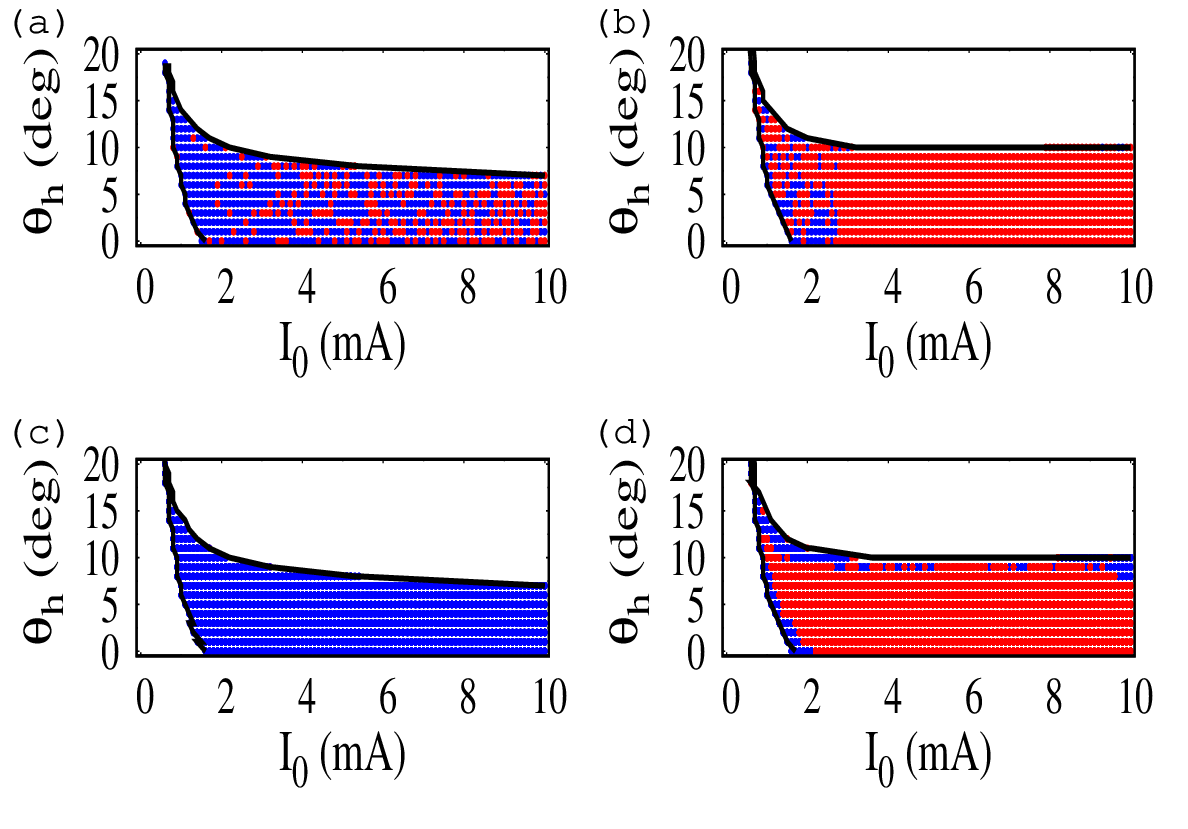}
	\caption{Two parameter diagrams between the field-angle $\theta_h$ and $I_0$ for (a) $N$ = 2, $\beta$ = 0, (b) $N$ = 2, $\beta$ = 0.2, (c) $N$ = 10, $\beta$ = 0 and (d) $N$ = 10, $\beta$ = 0.2 to indicate the regions of synchronized oscillations (red), desynchronized oscillations (blue) and no oscillations ( white).  The black colour line corresponds to the bifurcation boundary between nonoscillatory and  oscillatory regions.}
	\label{fig10a}
\end{figure}

 In this section, we consider specifically the impact of the field angle on synchronization of the array of STNOs.  For the case of $N$ small, we have plotted four two-parameter diagrams between the current $I_0$ and field-angle $\theta_h$ for (i) $N$ = 2, $\beta$ = 0, (ii) $N$ = 2, $\beta$ = 0.2, (iii) $N$ = 10, $\beta$ = 0 and (iv) $N$ = 10, $\beta$ = 0.2 in Figs.\ref{fig10a}(a), (b), (c) and (d), respectively. The red, blue and white colours refer to the regions of synchronized oscillations, desynchronized oscillations and no oscillations (steady states i.e. $\frac{d{\bf m}}{dt} = 0$ at $t\rightarrow \infty$), respectively.  
The regions of synchronzied ($R$ = 1) and desynchronized ($R\ne$ 1) oscillations are identified with the Kuramoto order parameter $R$. The no oscillatory region is plotted by verifying that the amplitudes of oscillations of the STNOs are zero. In the no oscillatory region the magnetizations of the entire set of STNOs settle at the same state. 
From Fig.\ref{fig10a}(a) we can understand that even in the absence of field-like torque and when $\theta_h$ = 0$^\circ$, synchronized oscillations are possible. However, the existence of synchronized oscillations is not continuous with respect to current even when the field-angle is varied. This clearly indicates that the change in the field-angle does not provide any impact on the synchronization. Fig.\ref{fig10a}(b) clearly implies that the presence of field-like torque makes the system to exhibit synchronized oscillations for a wide range of current. From Fig.\ref{fig10a}(b) we can observe the ineffectiveness of the field-angle over synchronization.  In the absence of field-like torque ($\beta$ = 0) even when the number of STNOs is increased from 2 to 10 the system does not show any synchronized oscillations until the field-angle is varied as shown in Fig.\ref{fig10a}(c). However, in the presence of field-like torque the system of 10 STNOs exhibit synchronized oscillations over a wide range of current and field-angle, see Fig.\ref{fig10a}(d). From Figs.\ref{fig10a}(a-d) we can understand that when the number of STNOs is increased even by a small number, the field-like torque is essential and the field-angle alone is ineffective for exhibiting synchronized oscillations.  Figures \ref{fig10a}(b) and (d), plotted for $N$ = 2 and 10, respectively imply that the field-angle reduces the value of current for the onset of synchronized oscillations.  We have drawn the bifurcation boundaries (black colour line) in Figs.\ref{fig10a}(a-d) between the regions of no oscillation and synchronized oscillations by determining the maximum transverse Lyapunov exponent (as discussed below in Sec.VIII). The value of maximum transverse Lyapunov exponent in the region bounded by the boundaries besides white region is below zero, which implies that the synchronized oscillations are stable. The system enters into the synchronized oscillatory region from no oscillatory region by Hopf bifurcation. The blue points corresponding to desynchronized oscillations in Figs.\ref{fig10a} corresponding to the numerical results within the boundary are due to the random choice of the initial conditions for computation.
\begin{figure}[htp]
	\centering\includegraphics[angle=0,width=1\linewidth]{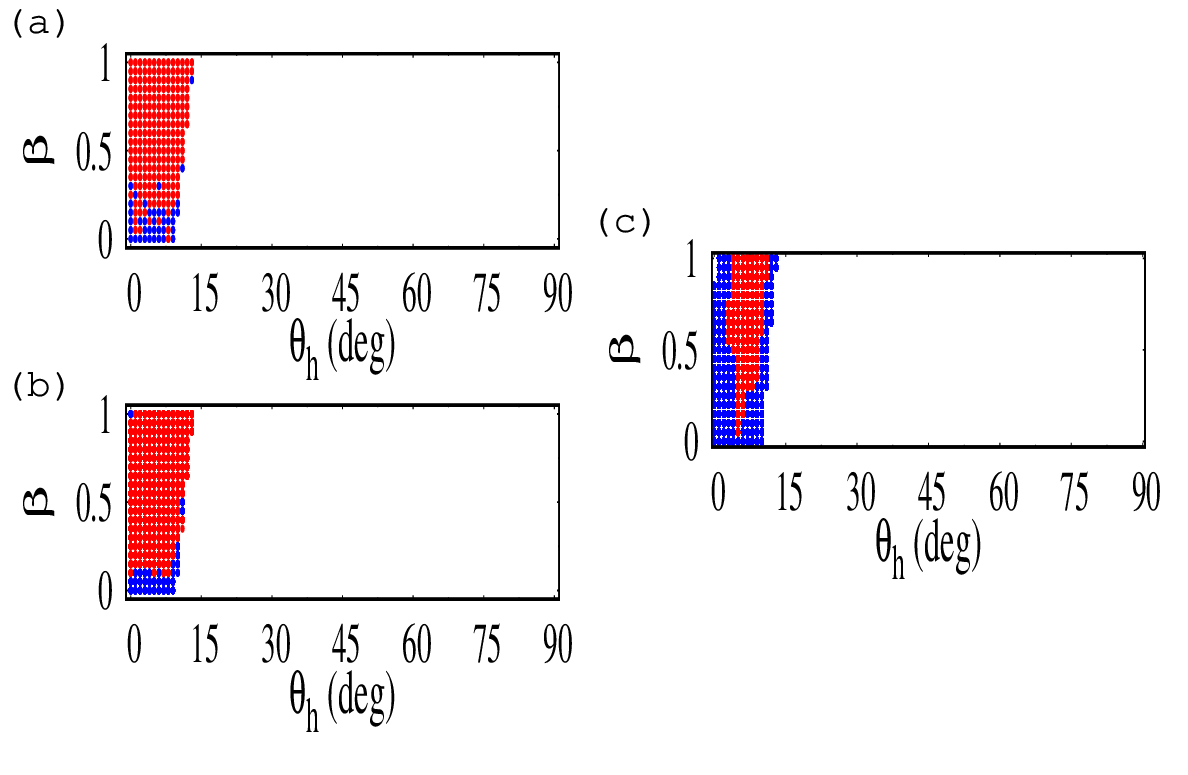}
	\caption{ Two parameter diagrams between the field-like torque $\beta$ and field-angle $\theta_h$ for (a) $N$ = 2 and (b) $N$ = 10 and (c) $N$ = 100 to identify the regions of synchronized oscillations (red), desynchronized oscillations (blue) and no oscillations ({\bf white}) when $I$ = 2.5 mA.}
	\label{fig10b}
\end{figure}
We have also plotted the phase diagram between $\beta$ and $\theta_h$ for $N$ = 2, 10 and 100 corresponding to the current 2.5 mA in Figs.\ref{fig10b}(a), (b) and (c), respectively, where the same colour code is maintained as in Fig.\ref{fig10a}. Figs.\ref{fig10b}(a) and (b) confirm that when the current is kept fixed, there is no impact of the field-angle on the synchronization and the field-like torque only plays a crucial role in making synchronization in the system of 2 and 10 STNOs. Fig.\ref{fig10b}(c) clearly shows that the change in the field-angle makes the system of 100 STNOs to oscillate synchronously.  

\section{Stability analysis of Synchronization of STNOs}
So far we have verified the existence of synchronized oscillations in arrays for 2, 10 and 100 STNOs by numerically computing the Kuramoto order parameter $R$. It quantifies the degree of their synchrony through the average phase. Further, from an experimental point of view, one of the main interests is to understand the conditions for the existence and stability of synchronization behaviour, as it can provide benefits in operating an array of STNOs to increase the overall radiative power \cite{tur1}. In this regard, we also adopt the most often applied tool, namely the largest Transversal Lyapunov Exponent (TLE) of the synchronization manifold \cite{pec}, which allows one to quantify the stability of synchronization. The analysis of TLE completely depends on the perturbation vector and its variational equation to calculate the most prominent Lyapunov exponent in the direction transversal to the synchronization manifold, so that it explicitly confirms the stability of the synchronous state(see below).
\begin{figure}[htp]
	\centering\includegraphics[angle=0,width=1\linewidth]{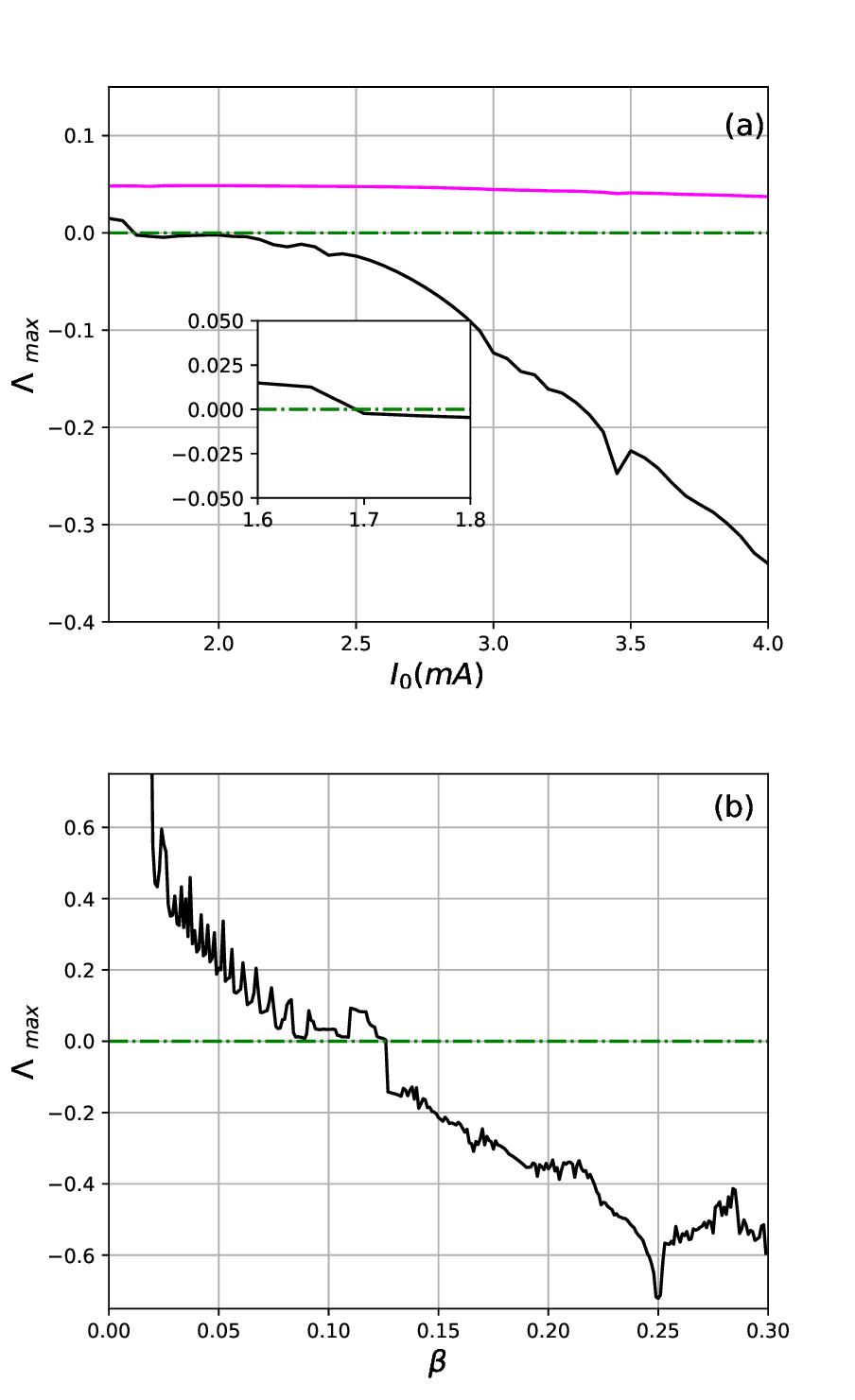}
	\caption{ Maximum transverse Lyapunov exponent for synchronized attractor of coupled STNOs obtained from the linearized equations (Eqs.(5)) versus  (a) current for  $\beta$ = 0 (magenta) and $\beta$ = 0.2 (black) and (b) field-like torque for $I_0$ = 4 mA. (Inset) Zoom-in view  of $\Lambda_{max}$  indicates that the value  $\Lambda_{max}$ becomes negative and the synchronous state becomes stable when the values of current are between 1.6 mA and 1.8 mA. Here $N$ = 2 and the value of the field angle is fixed as $\theta_h$ = 0$^\circ$.}
	\label{fig11}
\end{figure}
 In this section, we summarize our detailed analysis to check the stability of synchronization of large 1-D array of identical STNOs as $t\rightarrow\infty$  by finding the transverse Lyapunov exponents~\cite{tur1}. In this regard under a stereographic projection~\cite{lakprl,lak} Eq.(\ref{goveqn}) can be equivalently rewritten using the complex scalar function
\begin{align}
	z_{j}(t)=\frac{m_{j}^{x}+im_{j}^{y}}{1+m_{j}^{z}},~~~~j~=~1,2,3,....,N, \nonumber
\end{align} 
or equivalently
\begin{align}
	m_{j}^{x}=\frac{z_{j}+\bar{z}_{j}}{1+|z_{j}|^{2}},~~~~ m_{j}^{y}=-i\frac{z_{j}-\bar{z}_{j}}{1+|z_{j}|^{2}},~~~~ m_{j}^{z}=\frac{1-|z_{j}|^{2}}{1+|z_{j}|^{2}} \nonumber
\end{align}  
as 
\begin{align}
(i+\alpha)	&\frac{dz_{j}}{dt}=\frac{\gamma}{2}H_{a} sin \theta_{h}(1-z_{j}^{2}) - \gamma H_{a}cos \theta_{h} z_{j} \nonumber\\
	&-\frac{\gamma(H_{k}-4\pi M_{s})z_{j}(1-|z_{j}|^2)}{1+|z_{j}|^2}\nonumber \\&-\frac{\gamma}{2}\frac{ a(i+\beta)I_{dc}(1-z_{j}^2)}{1+\lambda\frac{z_{j}+\bar{z}_{j}}{1+|z_{j}|^2}} 
	\Bigg(1-\sum_{k=1}^{N}\beta_{\Delta R_{k}}\frac{z_{k}+\bar{z}_{k}}{1+|z_{k}|^2}\Bigg)^{-1}
	\label{complex} 
\end{align}

where  $a={\hbar \eta I_{dc}}/{2 M_s e V}$.

Rewriting the above equation in terms of real and imaginary variables $z_{j}=x_{j}+iy_{j}$, we get
\begin{subequations}
\label{complex1}
\begin{align}
&\frac{dx_j}{dt}=f(x_{j},y_{j})+ R_1(x_{j},y_{j}) \label{complex1a}\\
&\frac{dy_j}{dt}=g(x_{j},y_{j})+ R_2(x_{j},y_{j}) \label{complex1b}
\end{align}
\end{subequations}
where
\begin{align}
R_1(x_j,y_j) &= a Q(x_{j},y_{j}) \bigg[ 2(\beta-\alpha)x_{j}y_{j}-(1+\alpha \beta) (1-x_j^2+y_j^2)\bigg],\nonumber\\
R_2(x_j,y_j) &= a Q(x_{j},y_{j}) \bigg[ 2(1+\alpha\beta)x_{j}y_{j}+(\beta-\alpha) (1-x_j^2+y_j^2)\bigg],\nonumber
\end{align}
\begin{eqnarray}
	f(x,y)&=&\frac{\gamma}{1+\alpha^2}\Bigg\{\frac{H_a}{2} sin\theta_h [\alpha(1-x_j^2+y_j^2)-2x_{j}y_{j}]\nonumber \\ 
	&-& [H_a\cos\theta_h+H_k-4\pi M_s](\alpha x_{j}+y_{j}) \frac{1-r_j^2}{1+r_j^2} \Bigg \}, \nonumber\\
	g(x,y)&=&\frac{\gamma}{1+\alpha^2}\Bigg\{-\frac{H_a}{2} sin\theta_h [1-x_j^2+y_j^2+2\alpha x_{j}y_{j}]\nonumber \\ 
	&+& [H_a\cos\theta_h+H_k-4\pi M_s]( x_{j}-\alpha y_{j}) \frac{1-r_j^2}{1+r_j^2} \Bigg \}\nonumber,
\end{eqnarray}
where 
\begin{align}
	Q(x,y)=\frac{1+r_{j}^{2}}{1+r_{j}^{2}+2\lambda x_{j}}\bigg(1-\sum_{k=1}^{N}\beta_{\Delta R_k}\frac{  2 x_{k}}{1+r_{k}^{2}}\bigg)^{-1}, \nonumber
\end{align}
and
\begin{align}
	r_{j}^{2}=x_{j}^{2}+y_{j}^{2}. \nonumber
\end{align}
Further, Eqs.\eqref{complex1} are transformed to the transversal coordinates $x_{\perp}=u_{1}-u_{2}$, where $u_j=(x_j,y_j)$. Then they are linearized transverse to the synchronization manifold as
\begin{equation}
	\dot{x}_{\perp}=(J+K)x_{\perp},\label{linear}
\end{equation}
where $J$ is the Jacobian matrix of the vector field $F=(f(x,y),g(x,y))$ identified at the synchronization manifold, that is, $J=dF_(x_{s},y_{s})$, $x_{1}=x_{2}=...=x_{N}=x$ and $y_{1}=y_{2}=...=y_{N}=y$,  and $K$ represents the matrix that results from the linearization of the coupling terms ($R_1(x_j,y_j)$ and $R_2(x_j,y_j)$). The synchronized oscillations are said to be stable when $x_\perp\rightarrow 0$ as $t\rightarrow\infty$ or all of the transverse Lyapunov exponents ($\Lambda$) are negative.  The transverse Lyapunov exponents are the Lyapunov exponents associated with the linearized equations \eqref{linear}. The maximum  transverse Lyapunov exponent ($\Lambda_{max}$) for the case of 2 STNOs is plotted in Figs.\ref{fig11} for $\theta_h$ = 0$^\circ$. In Fig.\ref{fig11}(a) the value of $\Lambda_{max}$ is plotted for the absence and presence of field-like torque and from this figure we can identify that when the field-like torque is present the value of $\Lambda_{max}$ crosses from positive to negative value at $I_0$ = 1.7 mA (see the inset of Fig.\ref{fig11}(a)) and continues to be negative thereafter.  The negative value of the $\Lambda_{max}$ means that the coupled oscillators exhibit asymptotically stable synchronized oscillations. However, in the absence of the field-like torque the value of $\Lambda_{max}$ is always positive for the entire range of the current as shown in  Fig.\ref{fig11}(a). The negativity of $\Lambda_{max}$ for $\beta$ = 0.2 corresponds to the numerical results discussed in Fig.\ref{fig4}(a) where stable synchronized oscillations are obtained for $N$ = 2 above the value of  current $I_0$ = 1.7 mA when $\beta$ = 0.2 and $\theta_h$ = 0$^\circ$. This confirms the validity of the numerical simulations. In Fig.\ref{fig11}(b) the value of $\Lambda_{max}$ has been plotted against the strength of field-like torque for $I_0$ = 4 mA. We can observe that the $\Lambda_{max}$ becomes negative after $\beta\approx$ 0.13, which implies the possibility of stable synchronized oscillations by increasing the strength of the field-like torque and this corresponds to the results shown in Fig.\ref{fig3}(b).   The above analysis can be extended to more oscillators straightforwardly.
\section{Power}
\begin{figure}[htp]
	\centering\includegraphics[angle=0,width=1\linewidth]{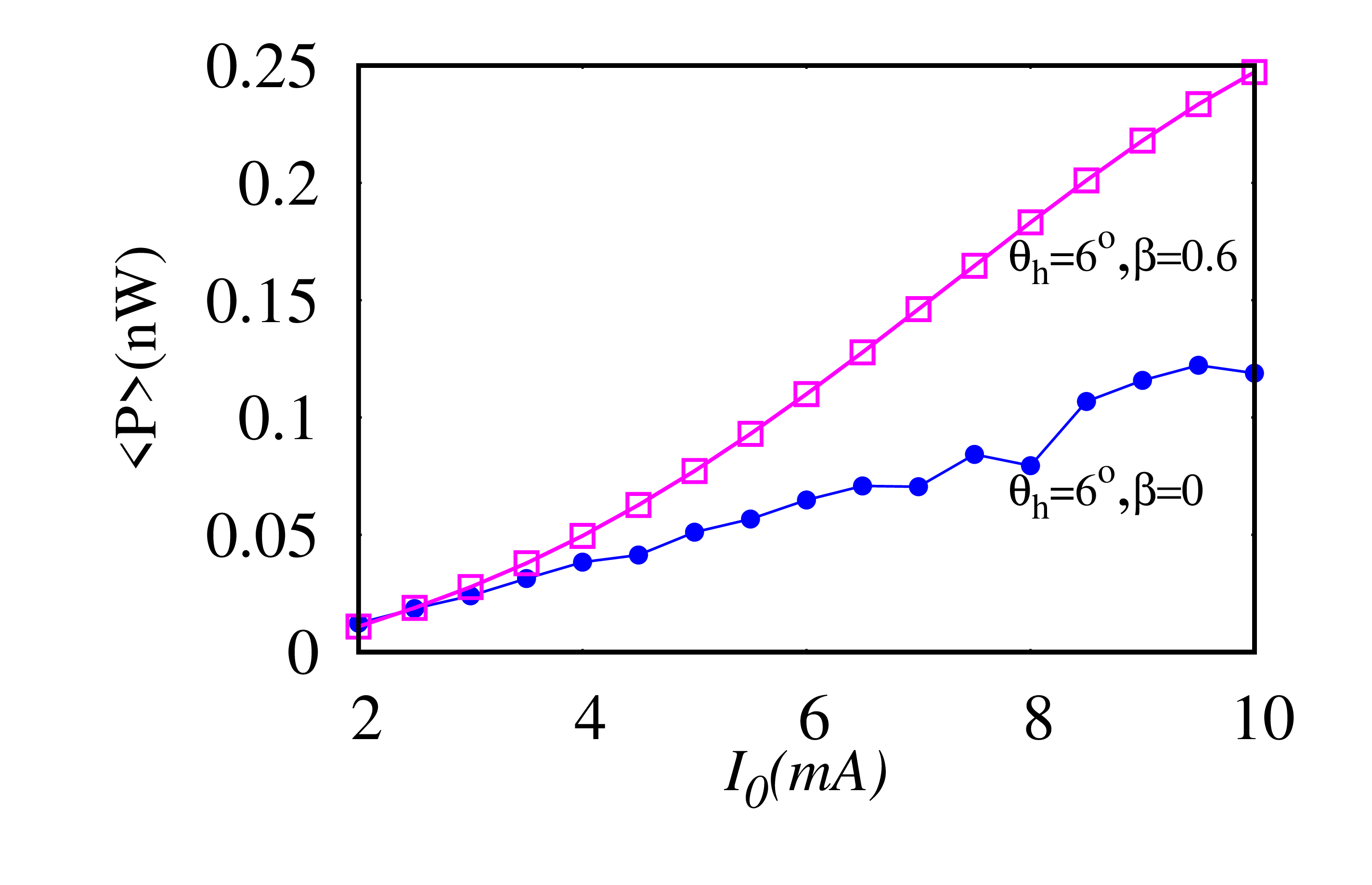}
	\caption{Average output power for a STNO in the desynchronized (blue line) state and synchronized (magenta) states. Here, the total array is with $N$ = 100.}
	\label{fig12}
\end{figure}
The output power $P$ corresponding to the output voltage $V$ of an STNO is given by~\cite{Russ}
\begin{align}
	P = \frac{V^2}{2R_C} = \frac{1}{2R_C}\left(\frac{I_0 ~\Delta R_i~ R_C~({\bf m}_i\cdot{\bf p}) }{R_{0i}+R_C}\right)^2 = \frac{I_0^2 ~{\Delta R_i}^2~ R_C~m_{xi}^2 }{2(R_{0i}+R_C)^2}. \label{S12}
\end{align}
The time averaged power is derived as~\cite{Russ}
\begin{align}
	<P> ~=~ \frac{I_0^2 ~{\Delta R_i}^2~ R_C }{2(R_{0i}+R_C)^2} ~<m_{xi}^2>_{time}. \label{S13}
\end{align}
$<P>$ is determined from $<m_{xi}^2>_{time}$ corresponding to the $m_{xi}(t)$ between 950 ns and 1000 ns. The power $<P>$ of an STNO in the array of 100 STNOs is computed when they are in the desynchronized state ($\beta$ = 0) and synchronized state ($\beta$ = 0.6) for $\theta_h$ = 6$^\circ$ and plotted against the current in Fig.\ref{fig12}.  It is observed that the power of an STNO can be enhanced well by the presence of field-like torque and this is attributed to the increase in the range of the oscillations.  At the current $I_0$ = 10 mA, the power is 0.247 nW, which is about two times larger than the power when $\beta$ = 0.  Furthermore, the power of $N$ STNOs in the synchronized state is $N^2$ times higher than that of a single STNO. This means that with the strength of the field-like torque $\beta$ = 0.6 and field-angle $\theta_h$ = 6$^\circ$ the power of 100 synchronized STNOs may reach about 2.47 $\mu$W which is quite desirable.

\section{conclusion}
We have observed the existence of complete synchronization in 2, 10 and 100 serially connected and electrically coupled STNOs with perpendicular magnetic anisotropy by incorporating the field-like torque and tuning the direction of the external magnetic field. The study is carried out by employing the macrospin simulation of the associated Landau-Lifshitz-Gilbert-Slonczewski equation.   It is observed that when the size of the system is small ($N$ = 2 or 10), the field-like torque is sufficient for achieving complete synchronization between the STNOs. On the other hand, for the system with a large 1-D array of STNOs ($N$ = 100), the direction of the external field is also needed to be tuned for a slight angle along the polarization of the pinned layer in addition to the presence of field-like torque to achieve complete and stable synchronization. The synchronized STNOs for the $N$ = 2, 10 and 100 cases exhibit large amplitude oscillations, which is beneficial for generating considerable output power. The order parameter is computed to check for synchronization ($R$ = 1) and desynchronization ($R \neq 1$). When the field direction's tunability or field-like torque's presence alone is considered, we identify cluster formation within the system. Also, when the field is tilted and fixed at an angle, the presence of the field-like torque forms clusters within the system, and its further increment in strength reduces the number of clusters, and eventually, complete synchronization is brought out by the field-like torque.

Furthermore, the increase in the strength of the field-like torque increases the range of the angle of the external field at which synchronization can be achieved. It is proved that the frequency of the completely synchronized oscillations can be tuned over a wide range of currents. The transverse Lyapunov exponent is calculated to confirm the existence of stable synchronization in coupled STNOs due to the field-like torque and to validate the results obtained in the numerical simulation.  The output power of a STNO is computed in the desynchronized state and verified that it increases more than two times in the synchronized state so that in an array there will be considerable increased power.  Finally, by considering the field-like torque and controlling the direction of the external magnetic field, one can bring a complete stable synchronization in a 1-D array of $N$ serially connected and electrically coupled STNOs with perpendicular anisotropy.  This analysis can also be extended up to a multi-layer array and network and we hope to pursue the study further in future.

\section*{Acknowledgements}
The works of V.K.C. and R. G are supported by the DST-SERB-CRG Grant No. CRG/2020/004353 and they wish to thank DST, New Delhi for computational facilities under the DST-FIST programme (SR/FST/PS-1/2020/135) to the Department of Physics.   M.L. wishes to
thank the Department of Science and Technology for the award of a DST-SERB National Science Chair  under Grant No. NSC/2020/00029 in which R. Arun is supported by a Research Associateship.\\~\\

\section*{DATA AVAILABILITY}
The data that support the findings of this study are available from the corresponding author upon reasonable request.

\end{document}